\documentclass[a4paper]{article}

\usepackage[utf8x]{inputenc}
\usepackage{titlesec}
\usepackage{url}
\usepackage{color}
\usepackage{authblk}
\usepackage{lineno,setspace}
\usepackage{amsmath,amssymb}
\usepackage{changepage}
\usepackage{textcomp,marvosym}

\usepackage{cite}
\usepackage{nameref,hyperref}
\usepackage{microtype}
\DisableLigatures[f]{encoding = *, family = * }
\usepackage[table]{xcolor}
\usepackage{array}

\raggedright
\usepackage[aboveskip=1pt,labelfont=bf,labelsep=period,justification=raggedright,singlelinecheck=off]{caption}

\makeatletter
\renewcommand{\@biblabel}[1]{\quad#1.}
\makeatother

\usepackage{lastpage,fancyhdr,graphicx}
\usepackage{epstopdf}

\newcommand{\blue}[1]{\textcolor{black}{#1}}

\title{Nine Quick Tips for Analyzing Network Data}
\author[1,*]{Vincent Miele}
\author[2]{Catherine Matias}
\author[3]{St\'ephane Robin}
\author[1]{St\'ephane Dray}
 
\affil[1]{Universit\'e de Lyon, F-69000 Lyon; Universit\'e Lyon 1; CNRS, UMR5558, 
Laboratoire de Biom\'etrie et Biologie \'Evolutive,
F-69622 Villeurbanne, France}
\affil[2]{Sorbonne Universit\'e,  Universit\'e Paris Diderot, Centre National de la Recherche Scientifique,  Laboratoire de Probabilit\'es, Statistique    et    Mod\'elisation,  F-75005 Paris, France}
\affil[3]{UMR MIA-Paris, AgroParisTech, INRA, Universit\'e Paris-Saclay, F-75005, Paris, France}
\affil[*]{Corresponding author: vincent.miele@univ-lyon1.fr}

\date{}

\begin{document}
\DeclareGraphicsExtensions{.pdf, .jpg, .jpeg, .png, .gif,.eps}
\maketitle
\begin{abstract}
  These tips provide a quick and concentrated guide for beginners in the analysis of network data. 
\end{abstract}

\section*{Introduction}

\blue{From the molecular to the ecosystem level, a biological system can often be represented as a set of entities that interact with other biological entities. Recent advances in data acquisition technology (e.g., high-throughput sequencing or tracking devices) open up the opportunity to quantify these interactions and call for the development of ambitious methodology to tackle these data. In this context,} networks are widely used in Biology, Bioinformatics, Ecology, Neuroscience or Epidemiology to represent interaction data \cite{ide17}. A network contains a set of entities (the {\it nodes} or {\it vertices}) that are connected by {\it edges} (or {\it links}) depicting some interactions or relationships. These relationships may be either directly observed or deduced from raw data. The first case encompasses protein-protein interaction (PPI) networks where interactions between two proteins are experimentally assessed or plant-pollinator interactions that are directly observed in the field. Gene regulatory networks reconstructed from gene expression data, co-occurrence networks inferred from species abundances or animal social contact networks deduced from GPS tracks are some examples of the second case. New kinds of networks are still emerging (for instance, cell-cell similarity networks\cite{zid18}, Hi-C networks and image similarity networks \cite{wan18}).

Networks are very attractive objects and many methods have been developed to analyze their structure. However, biological networks are often analyzed by non-specialists and it may be difficult for them to navigate through the plethora of concepts and available methods. In this paper, we propose nine tips to avoid common pitfalls and enhance the analysis of network data by biologists.

\titleformat{\section}
  {\normalfont\Large\bfseries}{Tip \thesection:}{1em}{}


\section{Formulate questions first, use networks later}\label{questions} 
Network theory is well established and truly powerful but it cannot be used as a ``black-box". Indeed, building a network should not be considered as an end in itself. 
We recommend to (i) establish a list of scientific questions and hypotheses before manipulating the data and then (ii) evaluate if these questions naturally translate into a series of network analyses, rather than making network analyses first and checking whether they raise questions after \blue{(in agreement with rule 1 in \cite{kas16})}. Indeed, it is generally immediate to represent/model the data with a network, but much trickier to translate a question into a network-based analysis.

To this end, \blue{besides integrating the network formalism, it is important to embrace the network viewpoint.} It relies on a cornerstone idea that makes the strength but also the challenge of network modelling: any interaction is considered within its context taking into account the other interactions that occur (or not). \blue{In this viewpoint,}  any interaction between the pair of nodes (A,B) \blue{is considered in the context of} the other pairs involving A or B. For instance, the importance of a particular edge between two genes will be differently assessed if the target gene is or is not a {\it hub} (i.e. regulated by many genes). 
This viewpoint does not consider interactions as independent objects and is thus the exact opposite of examining the set of interactions one by one. 

Finally, it is obviously recommended to check whether your questions and data really fit the network viewpoint before performing any analysis. 
\blue{If the number of nodes and/or edges is very low, network analysis can be applied but results can be disappointing as they are not enough observed interactions to identify a structure in the data. 
On the other hand, although any matrix can be viewed as a network (one edge per cell, see next Tip), it is often more adequate to consider using non-network methods dedicated to complete matrices. 
For instance a correlation matrix, possibly viewed as a correlation network, can be naturally analysed with a hierarchical clustering or a principal component analysis.}
In other words, network analysis is not necessarily the answer when analysing a data matrix. 

\section{Categorize your network data correctly}\label{categorize}
To grab the cutting-edge concepts and methods in the networks field, learning the appropriate vocabulary from {\it graph theory} is a prerequisite \cite{die16}.
In particular, it is important to categorize your network properly to be sure you apply suitable methods. Different network categories for different data lead to different approaches.

Links can be {\it directed} (from a {\it source} to a {\it target}), possibly including {\it self-loops} (e.g., a protein interacting with itself or cannibalism in food webs).
Ignoring this information for the sake of simplicity would actually betray the original data. 
When dealing with edges embedding a value (a {\it weight}), we strongly advise you to avoid transforming the network into a {\it binary} one using any {\it ad-hoc} threshold value. Indeed it clears a significant part of the available information \blue{because some aspects of the network structure might be undetected in the binarized network \cite{bar04}.}  
\blue{This binarization could be used as} an exploratory step \blue{only} (for instance, to facilitate a first visualization step - see Tip \ref{vizu}), \blue{but it can bias your analysis (e.g., a nested pattern can be observed in binarized ecological networks but no not in weighted ones \cite{sta13})}. 
Methods handling weighted networks are usually available and therefore more efficient. Furthermore, the data analyst must be very cautious since, in the literature, weights can be considered as intensity-based (the greater the weight, the stronger the edge is) as well as distance-based (the smaller the weight, the closer the nodes are).

Nodes can belong to different categories and edges can be allowed only between nodes of \blue{different categories} ({\it bi/tri/multi-partite} networks; e.g., nodes as hosts/parasites or as plant/fungus/seed dispersers \blue{\cite{pav18}}).
\blue{It is mandatory to select methods that handle this particularity.} For instance, many statistical approaches rely on the expected number of edges (\blue{e.g., in the computation of {\it modularity}}, see Tip \ref{metrics}) which is here clearly different compared to the \blue{unipartite} case. 

Finally, additional information on the nodes is often available. For instance, nodes can have spatial positions (e.g., nodes as habitat patches or farms in 2D, brain area in 3D) or can be associated to external \blue{{\it attributes} (e.g., species traits in a food web)}. 
\blue{This additional information can be explicitly considered in the analysis, either to understand if it contributes to organize the network \cite{mie14}, or to look for some remaining structure once accounted for its effect \blue{(e.g., spatial  \cite{exp11} or phylogenetic effect \cite{mar10})}. In the former case, a simpler but suboptimal alternative often consists in using this information {\it a posteriori} in the interpretation of results (e.g., explaining the structure of genetic networks with spatial information \cite{for09} or comparing network structure with {\it metadata} \cite{hri14}).}

\section{Use specific network analysis software}\label{tools} 
A range of \blue{versatile} software is dedicated to network analysis. It is therefore a waste of time trying to use unspecific tools. These software tools belong to two distinct categories that have pros/cons: graphical user interface (mouse-based navigation) and software packages (command line interface or programming). 
The first category is mainly dedicated to powerful and interactive visualization (see Tip \ref{vizu}). It includes the two major open source software tools \texttt{Gephi} and \texttt{Cytoscape}, both supported by an active community. They also offer the computation of some network metrics (\blue{the choice of a relevant metric is discussed in Tip \ref{metrics})}. The second category is dominated by the two leading general-purpose network packages \texttt{NetworkX} and \texttt{igraph}, but there exist plenty of more specific packages (for instance \texttt{bipartite} in \texttt{R}). Browser-based visualization \cite{ros15} recently emerged as an intermediate category, mostly based on a collection of \texttt{javascript} libraries (e.g., \texttt{Sigma.js}). 

That said, we strongly suggest that you learn programming and scripting your analysis \blue{(in agreement with papers in the ``10 simple rules" collection about computing skills and reproducibility \cite{san13,car18}).} \blue{Dealing with reproducible code enhances network research:} you can re-run with no effort the complete analysis on a modified version of your raw data, on different datasets and share the code with others colleagues interested in the modelling approach. Finally, there exist a limited set of common network file formats (e.g., adjacency list in the format \texttt{source target}) that you should adopt from the very beginning, \blue{ to easily switch between different software tools. }

\blue{Meanwhile, the data analyst should avoid a hasty use of the different functions implemented in these tools.
As underlined in Tips~\ref{metrics} and~\ref{clustering}, it is crucial to understand the metrics/methods before running functions, and to select the appropriate ones with respect to the questions and the data at hand.}

\section{Be aware that network visualization can be useful but possibly misleading}\label{vizu}
\blue{One powerful aspect of networks is their ability to depict complex data in a single object. It can be therefore tempting to represent networks graphically in two dimensions: nodes are spread in the plane and edges drawn with the objective to achieve the most aesthetic design (the nodes'positioning is called a {\it layout})}. This apparently simple task is in fact a very hard combinatorial problem. An active research community proposed a series of heuristics aiming at obtaining a nice network view in a reasonable time, despite the growing size of available networks. The aforementioned tools (see Tip \ref{tools}) embed a wide range of easy-to-use layouts. 

\begin{figure}[h!]
\centering
a) \includegraphics[width=0.2\textwidth]{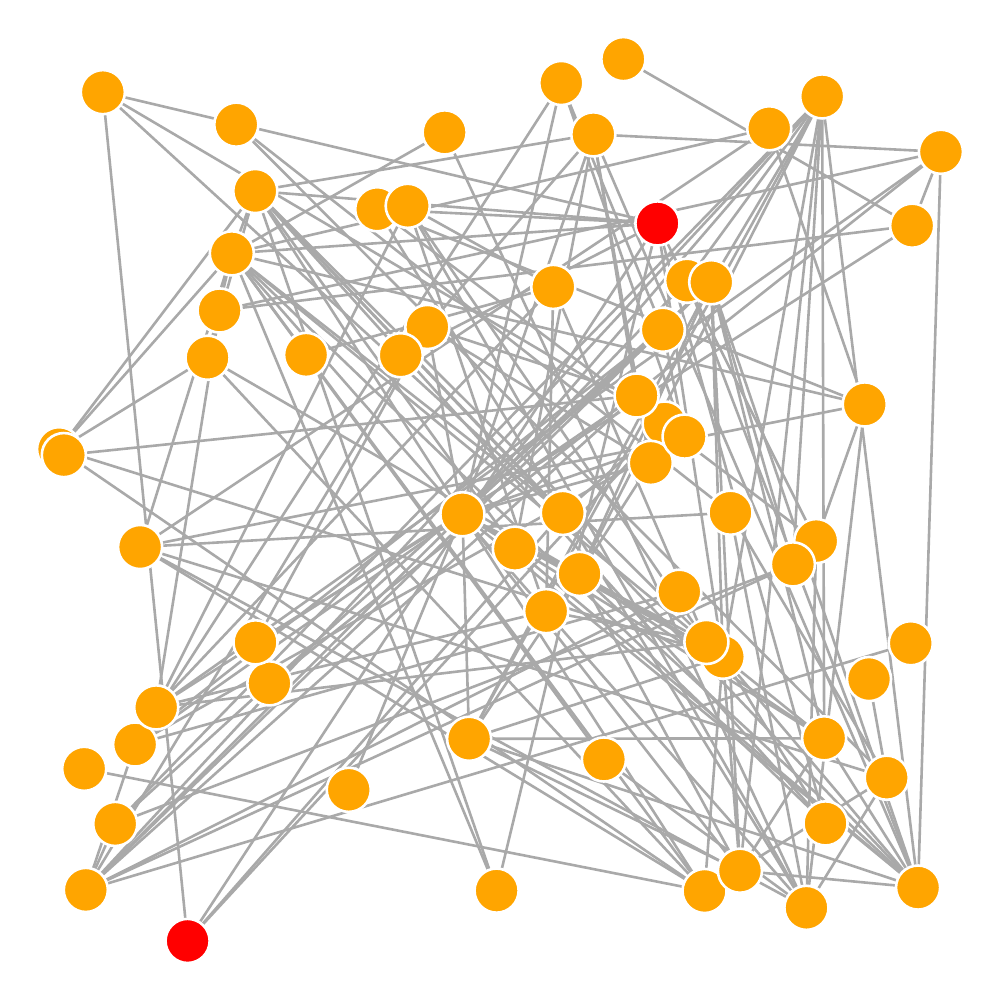}
b) \includegraphics[width=0.2\textwidth]{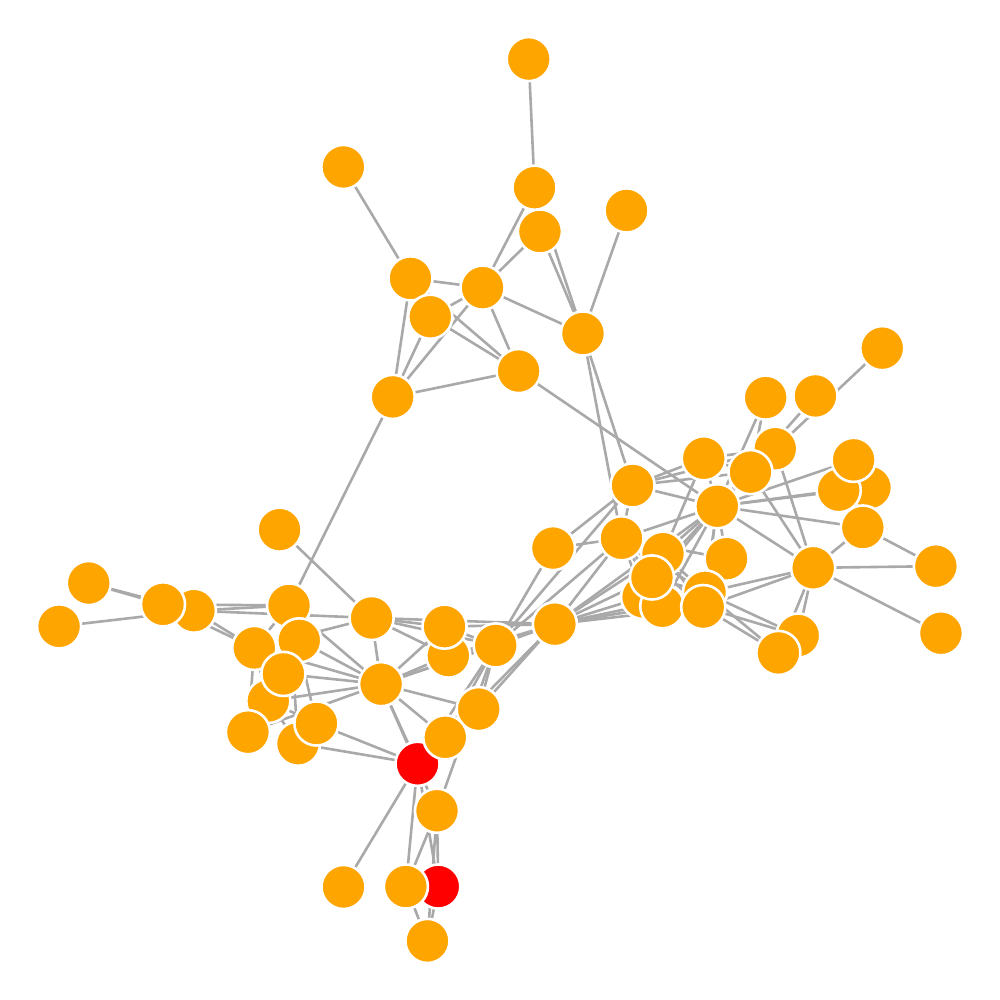}
c) \includegraphics[width=0.2\textwidth]{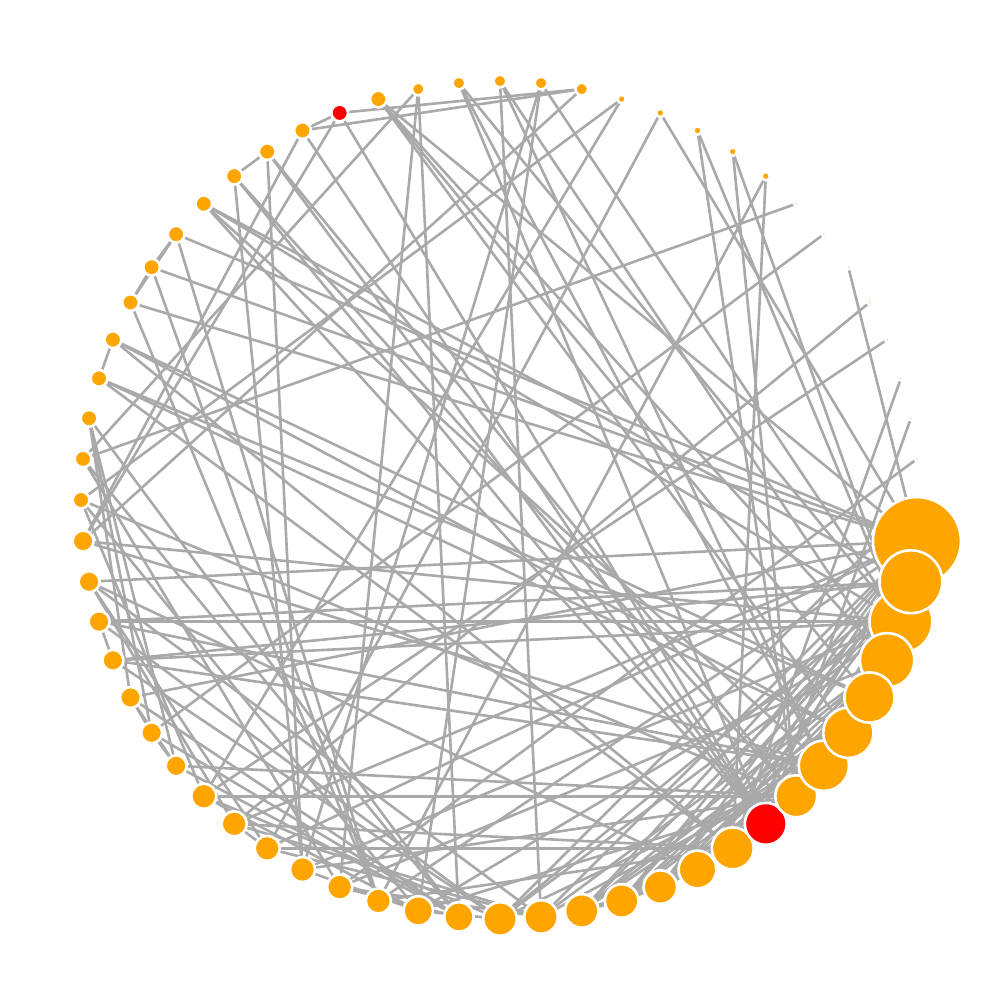}
d) \includegraphics[width=0.2\textwidth]{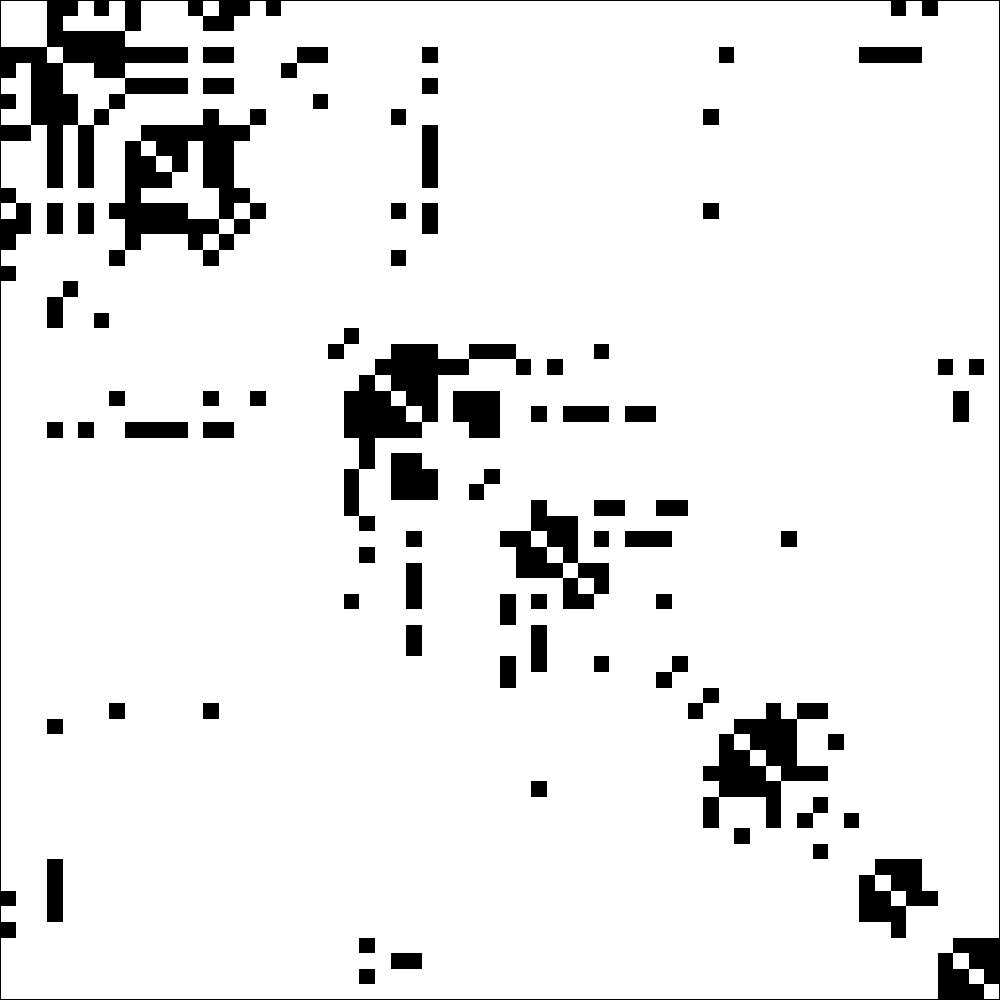}
\caption{\label{fig:vizu} 
\blue{
Four visualizations of the same network modelling interactions between 64 sociable weavers \cite{ros15,van14}.  a) Random layout. b) Fruchterman and Reingold layout.  c) Circle layout where nodes'size and position are defined by their degree. The same two nodes are colored in red in panels a-c to show their distance varies depending on the layout.  d) Representation of the adjacency matrix with row/columns ordering consistent with the clustering obtained with the Infomap algorithm (see \cite{for16} for details). Graphical representations are performed with the package \texttt{igraph}.}
}
\end{figure}

\blue{Graphics are usually considered as an important tool for exploratory data analysis \cite{tuk77}.} However, special care is required to not over-interpret network visualization. A layout does not only provide a nice representation of a network, it makes it optimal for a given set of objectives (e.g., maximizing attractions between connected nodes) that you often ignore. As a consequence, what you see with your eyes \blue{can be} biased. When visualizing a network, always keep in mind that the position of a node in such a display is not part of the data, but results from an algorithm. Hence, the distance between two nodes should not be interpreted as an intrinsic measure of proximity as another display algorithm would result in a possibly very different distance \blue{(see the distances between the two red nodes in Figure \ref{fig:vizu}a-b)}.
\blue{On the other hand, network visualization can be useful as a way to illustrate the results of a network analysis (as presented in Tips \ref{metrics} and \ref{clustering}). In this case, a layout should be chosen for its ability to highlight network properties (Figure \ref{fig:vizu}c) or conclusions drawn by an analysis (Figure \ref{fig:vizu}d). 
For instance, nodes can be positioned according to the values of some particular metrics of interest \cite{Krz11}. In any case, we encourage biologists to clearly describe the layout used in any graphical representation of a network in scientific publications, especially to make it reproducible.} 

Lastly, we also advise to consider visualizing the {\it adjacency matrix}  as a heatmap/a colored matrix \blue{(see Figure 1 in \cite{rub10} for an explanation)}. It allows to represent the presence or weight of edges (colored cells) but \blue{it has also the advantage to highlight edges' absence (blank matrix cells)}. This is particularly relevant when the matrix rows/columns are reordered in an informative manner (e.g., by increasing value of a metric \blue{\cite{bas03}} or according to some clustering results; see Tips \ref{metrics} and \ref{clustering} and Figure \ref{fig:vizu} d).

\section{Avoid blind use of metrics, 
\blue{understand}
formulas instead}\label{metrics}
\blue{Beside the limitations of network visualization, describing a network can also (and advantageously) consist in computing summary statistics}. The beginner will immediately find the path to a series of {\it network metrics}: one number per node or edge (\blue{local} metrics; e.g., {\it degree}) or one number for the whole network (\blue{global} metrics; e.g., {\it connectance/density} or modularity). Metrics have proliferated and it is strongly advised to take time to read carefully the mathematical definition of the metrics one has at hand (see also Tip \ref{rule:litterature}): the deeper the \blue{mathematical} understanding, the easier the interpretation is. 
\blue{For instance, the concept of {\it nodes'centrality} goes with a range of centrality metrics that have different meanings.}
\blue{Moreover, it is so easy to compute any metric with the aforementioned software tools that it can sometimes prevent the analyst to check their pros/cons.}
As an example, reading the definition of the widely-used {\it betweenness centrality}, you can understand it is based on {\it shortest paths}. If you intend to use this measure, it is therefore necessary to check whether the shortest path is a relevant concept associated to the process under study (such as energy fluxes in food webs) or if it is more questionable (\blue{e.g., paths in functional networks may not actually correspond to information flow \cite{rub10}, paths in contact networks may not be relevant when information or disease diffusion is not studied \cite{far15}}). 
\blue{Another example consists in the analysis of } directed and/or weighted networks with extensions of metrics to this case. 
It is important to note that the formula of the weighted degree accounts for two effects: how many {\it neighbours} and how large the weights are, two effects that are impossible to disentangle (a weighted degree of 2 can correspond to a single edge of weight 2 or four edges of weights 0.5). A similar problem can also be raised for the weighted path \blue{(potential pitfalls highlighted in \cite{cos18})}.
Lastly, global metrics are often used to compare networks (networks measured from different data or conditions, or simulated networks as mentioned in Tip \ref{simulation}). \blue{In this case, special care should be taken when comparing values because metrics differences can be a side effect of differences in simple network characteristics such as the number of nodes or edges (see common pitfalls mentioned in \cite{van10} for brain networks and a discussion on co-variation of metrics with characteristics of ecological networks in \cite{pel18}). For instance, modularity, number of modules and network size are known to be intertwined \cite{for07}}.

\blue{It is not unusual that authors, instead of choosing a given metric adapted to a particular question, compute a high number of metrics among the available ones.} However, many metrics are \blue{correlated (see a correlation study in \cite{far15}) and it becomes necessary to deal with this redundancy to interpret the results (e.g., with an ordination method \cite{kor19})}. This approach is not hypothesis-driven as recommended in Tip \ref{questions} and can undeniably be replaced by an incremental approach where metrics are selected one at a time for their ability to check particular hypothesis associated to the fundamental questions on the data \blue{(as for many statistical analysis, see rule 5 in \cite{kas16})}.

\section{
	Avoid blind use of clustering methods, check their difference instead}\label{clustering}
\blue{With the data avalanche arising this decade, leading to larger networks, clustering has become one of the most popular tool to get a comprehensive view of the network structure. Its general purpose is to aggregate nodes into {\it clusters} in order to identify a {\it meso-scale} structure in the network (i.e. zooming out the network). Choosing a network clustering raises similar issues than for choosing a network metric (Tip \ref{metrics}). It is much more than using one of the functions available in a software. Indeed, likewise any clustering methods, the ones dedicated to networks aim at gathering similar objects (i.e. nodes) and thus relies on a specific definition of node similarity. What does the analyst want to be similar in a network? Discussing the pros/cons of the different methods is beyond the scope of this article whereas a massive literature on the topic exists (see Tip \ref{rule:litterature}). However, we illustrate the impact of choosing a specific definition for node similarity with three classical proposals (among others).}

\blue{A first and natural definition for the similarity between nodes is the existence of a connection between them. Based on this definition, network clustering consists in finding a {\it modular} structure, i.e. identifying dense clusters  of nodes (also called {\it modules} or {\it communities}) poorly connected with others. {\it Community detection} methods \cite{for16} implement this approach, which implicitly assumes the existence of modules in the network. \blue{They were successfully applied in many studies in Biology (for instance to identify chromatin domains \cite{nor18}).} 
A second approach considers that two nodes are similar when they tend to be connected (or unconnected) with the same type of nodes. Hence, species in a food web are considered similar if they have similar preys and predators \cite{all09}. 
This definition can accommodate networks with non-modular structure \cite{new07} since it assumes that the nodes are involved in a ``diversity of meso-scale architectures" \cite{bet18}. 
The {\it stochastic block model} (SBM) is a popular method based on this definition \cite{new07,dau08}, which has shown to be relevant for the analysis of some biological networks (to highlight the complex architecture of connectomes \cite{bet18} or functional groups in ecological networks \cite{kef16}). One important feature is that it allows to model explicitly edge directions and weights by means of different statistical distributions \cite{mar10}. 
A third approach consists in associating a vector of characteristics to each node and then to gather nodes with similar characteristics. This includes {\it motifs}-based approaches\cite{sto12} and a wide range of innovative {\it node's embedding} techniques \cite{per14,gro16}. Nodes are described as points in a space with reasonable low dimension, which allows to apply the huge variety of existing clustering methods for  multivariate data. It is important to realize that each of these similarity concepts naturally results in different nodes clustering. The choice between these alternatives must be driven by biological questions not by their availability in software tools (Tip \ref{questions}).}

\section{Don't choose the easy way when simulating networks}\label{simulation}
\blue{
To highlight the properties specific to an observed network (for instance a peculiar metric value), a common practice consists in comparing with simulated networks.  These properties are detected as {\it significant} deviation (or not) from a {\it typical} behaviour implemented in simulated networks. However, there is no generic definition of a typical network and, as a consequence, the features that can be detected depends dramatically on the {\it null model} used to simulate networks. This null model must be chosen for a given purpose, fitting expected behaviours whereas contrasting those we are interested in.} \blue{In other words, it must fit the data reasonably well to avoid numerous false discoveries, but not too well so that deviations can emerge.}

A natural option could consist in selecting a null model among the series of {\it random graph models} (e.g., {\it Erdös-Rényi, small-world, scale-free, SBM, Exponential Random Graph, configuration model}). However, we recommend not to use them too hastily because they are often too general. \blue{For example, the Erdös-Rényi model (all edges independent and having the same probability of occurrence) is often a poor null model to detect nodes having an unexpectedly high degree. Indeed, it induces a Poisson degree distribution which is so far from the one observed in most networks that many nodes appear to be unexpectedly connected. On the opposite, no node can display an unexpectedly high degree with respect to the configuration model, as this null model precisely fits to the degree of each node.}
Moreover, the analyst is usually aware of a series of properties that should be displayed by a simulated network: imbalanced degree distribution, different nodes' roles associated with available side information, forbidden interactions \blue{(e.g., depending on body-mass in food webs \cite{bro06})}, etc. 
Such expected properties must be encoded in the simulation process \blue{(for instance a fixed degree sequence \cite{kef16})}, otherwise they will emerge and \blue{be detected as significant, or contribute to detect false significant effects as side effects.}
As an example, when assessing whether the number of feed-forward loops is unexpected in a given transcription network, the simulation procedure must rely on fixed number of nodes and degrees whereas the number of these loops remains free. 

Lastly, when the network under study is not directly observed but built from raw data interpretation, it can be relevant to simulate the whole construction process. 
Consider the case of contact networks inferred from movement data \cite{far15}: one can either simulate trajectories keeping some properties of the original data and then build a contact network, or directly simulate a ``realistic" contact network. The former approach will intrinsically account for the uncertainties and biases induced by the construction steps, which are likely to be overlooked by the later approach.

\section{Reconsider the data to build multiple network layers}\label{layers}
A network \blue{object} can be the result of data aggregation. 
Indeed, interactions are often observed at different times, locations or for different conditions. 
You are therefore strongly urged to keep in mind (and at hand) the different {\it layers} of data (time, space, type,...) and consider networks composed by multiple layers, because {\it multilayer networks} can provide new insights compared to an aggregated one \cite{boc14,pil17,bia18}. 

A network is called {\it dynamic} when it gathers a time series of network snapshots corresponding to successive rounds of data collection (the nodes' list possibly varying in time). \blue{In this case, the temporal variability of the network structure  can be assessed (e.g., rewiring of interactions or changes in network metrics over time) and extensions of the concepts developed in Tip \ref{clustering} now exist in the dynamic case \cite{mat17,ros18}.} 
\blue{For instance, the dynamics of animal social structure can be inferred from dynamic networks to enhance the understanding of disease transmission \cite{far17}. On another hand, interactions can be observed at different spatial locations. In ecology, they are often aggregated in a {\it metanetwork} (or {\it metaweb} \cite{ohl19}) to study how the local networks differ from this metanetwork and explain these variations with environmental factors. In these two cases, multiple layers allows to describe a network as an evolving object and the analysis aim to identify the spatio-temporal variations of interactions and their drivers.}

\blue{Different kind of interactions can also be observed between nodes. Stacking layers representing molecular interactions in different human tissues \cite{zit17} or mapping extrasynaptic and synaptic connectomes \cite{ben16} leads to a {\it multiplex} network: between any two nodes, there possibly exist more than one edge, one per interaction type at most (often visualized  with different colors). 
Taking jointly into account the different layers enhances the understanding of the nodes' interplay.}
\blue{For instance, using jointly trophic and non-trophic interactions enhances the definition of species ecological roles compared to the use of single layers independently \cite{kef16}. Finally, it is also possible to integrate different layers of information with different sets of nodes for each layer, such as proteins and chemical compounds \cite{ber16}. In this case, different kind of interactions are defined inside and between layers. In all these cases, different information layers are integrated into a comprehensive network such that they are treated jointly rather than one after the other.}

\section{Dive into the network literature, beyond your discipline}\label{rule:litterature}
Network science has emerged \blue{``at the dawn of the 21st century" \cite{bar19}}. It now involves a hyper-active community of researchers from different domains such as Physics, Statistics, Computer Science or Social Science. As a result, a massive literature on networks exists and it is challenging for biologists to dive into it. Indeed, we are not used to explore the bibliography outside our research domain. \blue{Reference books \cite{die16,new18,bia18,bar19} and reviews \cite{for16,boc14,goy18}}  are obviously good entry-points for developing your network skills. However, without any doubt you will highly benefit from a round trip in this literature exogenous to your field, provided that you make the effort to learn the appropriate vocabulary of this area.

\section*{Conclusion}
The 9 tips presented here should be a way for the data analyst to get a foot in the door of network data analysis. These tips are not exclusive and we are aware of other network-based questions that deserve a special interest, including diffusion on networks for instance. Still, the network non-specialist must be confident in his ability to learn, step by step, the network concepts and methods with a productive effect on his scientific questions.

\section*{Acknowledgements}
This work was partially supported by the grant ANR-18-CE02-0010-01 of the French National Research Agency ANR (project EcoNet).

\bibliographystyle{ieeetr} 
\bibliography{main}
\end{document}